\documentclass[amsmath,amssymb,superscriptaddress,prb,twocolumn]{revtex4}
\usepackage{graphicx}

\begin{document}

\title{
Finite size melting of spherical solid-liquid aluminium interfaces}

\author{Johan Chang$^{1,2}$, Erik Johnson$^{2,3}$, Takamichi Sakai$^{1}$ and Hiroyasu Saka$^{1}$ }
\affiliation{$^{1}$Department of Quantum Engineering, Nagoya University, Nagoya 464-8603, Japan\\
$^{2}$Niels Bohr Institute, Universitetsparken 5, DK-2100 Copenhagen, Denmark\\
$^{3}$Department of Materials Research, Ris\o\ National Laboratory, DK-4000 Roskilde, Denmark}
\date{\today}

\begin{abstract}
We have investigated the melting of  nano-sized cone shaped
aluminium needles coated with amorphous carbon using transmission
electron microscopy. The interface between solid and liquid aluminium
was found to have spherical topology. For needles with fixed 
apex angle, the depressed melting
temperature of this spherical interface, with radius $R$, was found to scale linearly with the 
inverse radius $1/R$. However, by varying the apex angle of the needles we show that
the proportionality constant between the depressed melting temperature and the inverse
radius changes significantly. This lead us to the conclusion that the depressed
melting temperature is not controlled solely by the inverse radius $1/R$.
Instead we found a direct relation between the depressed melting temperature and the ratio 
between the solid-liquid interface area and the molten volume.

\end{abstract}

\maketitle
\section{Introduction}
Thermodynamic properties of a material with reduced dimensionality can be markedly different from the bulk properties.\cite{RPP,RMP}
The melting temperature $T_c$ of materials confined to nano-meter
size can, for example, be significantly different from the bulk 
melting temperature $T_m$.
Since the first experimental evidence,\cite{takagi}
melting of nano-scale metal clusters has been a
case study~\cite{buffat,berman,allen1986,xray,xray1,dippel,lai,lai1,zhang,breaux,imura,johnson,cantor,ohashi,unruh}, involving a number of experimental techniques such as
transmission electron microscopy (TEM),\cite{buffat,berman,allen1986}
X-ray diffraction, \cite{xray,xray1} scanning tunneling microscopy (STM),\cite{dippel} and calorimetric
methods.\cite{lai,lai1,zhang,breaux}
For a recent review on size dependent melting see Ref.~\onlinecite{mei2007}.
 For free standing particles 
a depressed
 melting temperature $\epsilon=(T_m-T_c)/T_m$  proportional to the inverse radius $1/R$ is typically observed.
 However, a
 number of new investigations have reported a possibility of a
 non-linear
 relationship between $\epsilon$ and $1/R$ when the 
particles are sufficiently small.\cite{allen1986,nanda,dippel}  
These contradictory results give rise to the fundamental question:
what is really controlling the depressed melting temperature?
Several phenomenological models address the question of size dependent
melting temperature.\cite{pawlow,hanszen,chang1,sun,sunREV,delogu} The problem is complicated
by the fact that different crystal faces have different surface
energies and by the occurrence of surface premelting~\cite{lipowsky,pluis,kofman,kofmanrc} that depends on the surface
energies.
 In this paper we present an experimental sample geometry that, as we will see, avoids
 these complications.

 We use transmission electron microscopy (TEM) to
 investigate the size dependent melting point of nano-sized cone shaped  aluminium 
 needles as a function of the apex angle.
The interface between solid and liquid aluminium was found to have a spherical topology thus
the problem of crystal facets is eliminated. 
For a needle with fixed apex angle $\alpha$ we find a linear relation between
the depressed melting temperature $\epsilon(T_c)$ and the inverse radius $1/R$ of the spherical 
solid-liquid interface
similar to the observations for spherical particles where $\epsilon(T_c)=\Omega/R$ 
in which case $\Omega$ is a size-independent constant. 
However, by varying the apex angle we show that 
the proportionality constant $\Omega$ 
changes significantly. This implies that the depressed melting is not solely controlled
by the inverse radius. Instead, we argue that the depressed melting 
is determined by the ratio between the solid-liquid interface area and the molten volume. 

 \begin{figure*}
 \begin{center}
\includegraphics[width=0.95\textwidth]{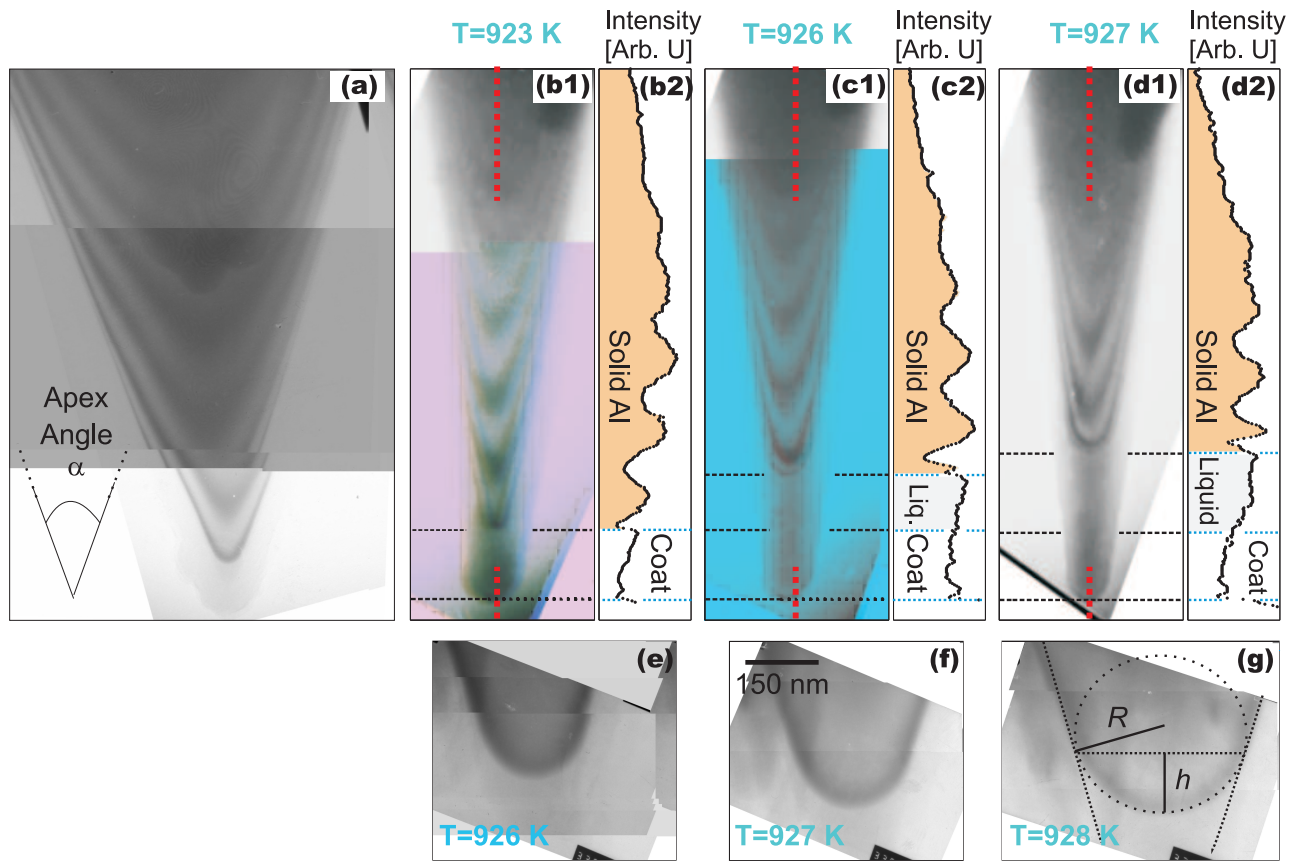}
\caption{ (a)-(b1) Bright-field images of needles with apex angle $\alpha=40^{\circ}$ and $26^{\circ}$, respectively.
The definition of the apex angle is shown schematically in the bottom left of (a).
 (b1)-(d1) Bright-field images of the aluminium needle with apex angle $\alpha=26^{\circ}$, for three different temperatures $T=923, 926$ and 927~K, respectively. In (b1) no melting is observed 
while in (c1)-(d1) a spherical melting front is visible. (b2)-(d2) Corresponding intensity profiles 
along the symmetry axis of the needle. (e)-(g) Similar needle recorded at similar temperatures
but magnified six times compared to (b1)-(d1). }\label{fig1}
\end{center}
\end{figure*}

\section{Methods} 
Samples with cone shaped aluminium needles were fabricated, from 99.99$\%$ pure
 aluminium, using the so-called ion digging method.\cite{iondigging} A
 suspension of fine diamond powder was dispersed on the vertical edge of a 0.1mm thin Al
 semi-disk (3mm diameter). The samples were mounted in an ion milling machine 
 equipped with a nitrogen cooled stage. 
 Sputtering with argon ions in the disk plane results in
 formation of thin needles at areas initially protected by diamond
 particles that were eventually 
 sputtered completely away. With this technique cone shaped Al needles with apex angles of $\sim$20-50
 degrees and diameters at the needle point around $\sim$20-30
 nm could be fabricated, see Fig.~\ref{fig1}(a) and (b1). A schematic illustration of this lithographic mask technique
 together with some preliminary results on Sn needles 
 were given in Ref.~\onlinecite{chang}.

In order to avoid evaporation, of aluminium, during the {\it in situ} heating experiment
 the whole surface of the needle samples were coated with a 50-80 nm
 thick amorphous carbon layer in a plasma coating apparatus.\cite{kato} In
 addition to protecting the needles, the carbon coating also helps to maintain the cone shape of the liquid
 phase during the {\it in situ} experiment.  
All samples were examined in a JEOL 200CX transmission electron microscope (TEM) using an operating voltage of 200kV and the {\it in situ} melting experiment was performed by making use of a side-entry heating holder. The temperature was recorded with a thermocouple (Pt-PtRh 13$\%$) placed in contact with the specimen cartridge.
The errors that could occur  in the temperature measurements, due to for example electron irradiation, are systematically small and do not disturb the real physical tendencies.\cite{senda,chang}
Conventional bright-field and dark-field imaging techniques were used to observe the solid-liquid interface.

\section{Results}
In this paper, we are going to study the melting process of four Al needles with apex angles $\alpha=$26, 29, 36 and 42 degrees. 
However, we start by inspecting the TEM micrographs of needles with $\alpha=40^{\circ}$ and $\alpha=26^{\circ}$ recorded at $T=923$~K before the melting initiates at $T=T_i$, see Fig.~\ref{fig1}(a) 
and (b1).
Both needles display a high degree of symmetry which is further supported by 
the symmetry of the thickness fringes. These TEM images therefore demonstrate that the needles
have an almost perfect conical shape.
To inspect the images more carefully it is useful to plot the intensity profile 
 along the central axis of the needle as indicated by the vertical dashed lines in Fig.~\ref{fig1}(b1).
Moving from the thick part of the needle (top of Fig.~\ref{fig1}(b1)) 
towards the needle tip, the intensity displays a period oscillatory behavior (Fig.~\ref{fig1}(b2))
that stems from the thickness that varies along the needle.
Eventually the intensity displays a discontinuity at the point between
the needle tip and the carbon coat as indicated by the top horizontal dashed line in Fig.~\ref{fig1}(b1) and (b2). 
Finally, the intensity varies slowly without any 
oscillations along the carbon coat due to its amorphous structure, see  Fig.~\ref{fig1}(b2).

Now we turn to discuss the finite size melting of the needles in the temperature 
range $T_i<T<T_m$. In Fig.~\ref{fig1}(c1) and (d1)
we show the same needle as in Fig.~\ref{fig1}(b1) but recorded at $T=926$~K and $T=927$~K,
respectively. In both cases the tip of the needle is molten. 
This fact is also  visualized by the intensity profiles in Fig.~\ref{fig1}(c2) and (d2).
The discontinuity between the needle tip and the carbon coat is no longer visible and another 
discontinuity appears at a thicker part of the  needle, see Fig.~\ref{fig1}(c2) and (d2).
We interpret this in terms of a melting process where the top part of the needle is molten 
and the discontinuity of the intensity profile now marks the interface between solid and liquid 
aluminium. As neither the liquid aluminium nor the amorphous carbon coat have any crystal 
structure we can not easily distinguish  these two elements from our bright-field images.
This explains why the discontinuity that appears at the needle tip for $T<T_i$ disappears
for $T>T_i$.
\begin{figure}
\includegraphics[width=0.45\textwidth]{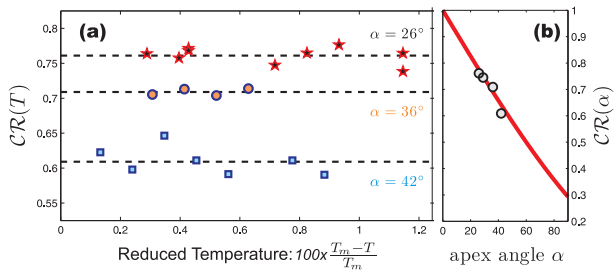}
\caption{\label{fig:fig3} (a) Temperature dependence of the cap ratio $\mathcal{CR}(T)$ for the needles with apex angles $\alpha=26^{\circ}$, $\alpha=36^{\circ}$, and $\alpha=42^{\circ}$. The dashed lines are fits to a constant. (b) The cap ratio $\mathcal{CR}(\alpha)$ as a function of the apex angle $\alpha$. The dashed line is explained in the text.}
\end{figure}

Next, we are going to discuss the topology of the solid-liquid interface.
A careful inspection of the image shown in Fig.~\ref{fig1}(c) reveals 
that the first dark fringe, in contrast to the other dark thickness fringes,
has a circular shape. This fringe indicates the position of 
the solid-liquid interface and it strongly suggests that it 
has a spherical topology. 
This was further 
confirmed by bright-field images taken with six times higher 
magnification shown in Fig. \ref{fig1}(e)-(g).
As the temperature increases from $T=926$~K in Fig. \ref{fig1}(e)
to $T=928$~K in Fig. \ref{fig1}(g), the solid-liquid interface area 
becomes larger as the melting front moves towards the thicker part of the 
needle.  However, the solid-liquid interface remains spherical at all
the observed temperatures.
Finally we emphasize that during alternating heating and cooling cycles no hysteresis was observed between melting and solidification -- indicating that the system is in a true equilibrium at all temperatures. Furthermore, the system does not seem to favor surface premelting since no precursor of the melting could be detected.  
 
 We are now going to describe the interface between the solid and liquid phases in greater detail.
 Since the solid-liquid interface forms a spherical cap at each temperature $T_i<T<T_m$, 
 one can  extract a cap height $h(T)$ and a spherical radius
 $R(T)$ directly from the TEM micrographs as shown in Fig.~\ref{fig1}(g).
 In Fig.~\ref{fig:fig3}(a) we show the  cap ratio $\mathcal{CR}(T)=h(T)/R(T)$
 as a function of the reduced temperature $\epsilon(T)=(T_m-T)/T_m$ for the needles with $\alpha=26^{\circ},36^{\circ}$, 
 and 42$^{\circ}$.
   This reveals the following important facts:
   {\it (i)} the cap ratio $\mathcal{CR}$
   is essentially temperature independent for each of the needles
   and {\it (ii)} the cap ratio $\mathcal{CR}$ decreases systematically with increasing apex angle 
   $\alpha$.
   This implies that the cap geometry remains unchanged as the melting 
   front moves towards the thicker part of a needle with fixed $\alpha$.
   However, the  cap ratio $\mathcal{CR}$ varies systematically as a function of 
   apex angle $\alpha$. 
   For each of the needles  $\mathcal{CR}(\alpha)$ was 
   extracted
 by constant fits to the data, see dashed lines in Fig.~\ref{fig:fig3}(a).
 The values of $\mathcal{CR}(\alpha)$, given in table~\ref{tab:tab1}, are shown in Fig.~\ref{fig:fig3}(b).
 
     \begin{figure}
\includegraphics[width=0.5\textwidth]{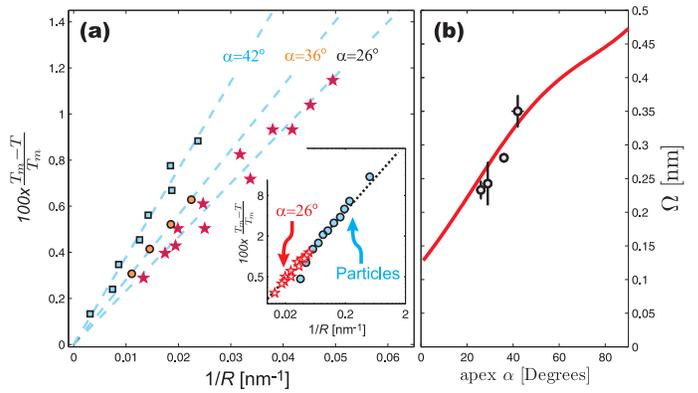}
\caption{\label{fig:melt} (a) The depressed melting temperature $\epsilon(T_c)=(T_m-T_c)/T_m$ as a function of the inverse radius $1/R$ for  needles with three different apex angles $\alpha=26^{\circ},~36^{\circ}$, and 42$^{\circ}$. The insert compares, in double logarithmic scales, the depressed melting temperature $\epsilon(T_c)$ of a needle with $\alpha=26^{\circ}$ and aluminium particles on a substrate (from Ref.~\onlinecite{lai1}).  (b) The melting rate $\Omega(\alpha)$ as a function of apex angle $\alpha$. The solid line is explained in the text.}
\end{figure}
   
   We are now going to describe the size dependent melting of the  aluminium needles on a more quantitative level.
 Once the melting process is initiated for $T^{\prime}>T_i$ a spherical solid-liquid interface 
 with a radius $R^{\prime}$ is formed. 
 Increasing (decreasing) the temperature $T^{\prime}$ leads to a corresponding increase (decrease) in
 the radius $R^{\prime}$ of the spherical interface. Thus the temperature $T^{\prime}$ associated
 with $R^{\prime}$ is in fact the melting temperature $T_c(R)$ of the spherical solid-liquid interface.
 By controlling the temperature $T$ and measuring $R$ on the TEM micrographs, it is
 possible to track the size dependent melting temperature $T_c(R)$ as a function of $R$.   
 Figure~\ref{fig:melt} displays the reduced melting 
 temperature $\epsilon(T_c)=(T_m-T_c)/T_m$ as a function of the inverse radius 
 $1/R$.
 As for spherical metal particles there seems to be a linear relation between the depressed
 melting temperature $\epsilon(T_c)$ and the inverse radius $1/R$, {\it i.e.}, 
 \begin{equation}
 \epsilon(T_c)=\frac{\Omega(\alpha)}{R(T_c)}
 \end{equation}
where $\Omega(\alpha)$ is a constant that depends on the apex angle $\alpha$. 
For a comparison, the inset of Fig.~\ref{fig:melt}(a) shows the depressed melting temperature 
of aluminium particles supported by a substrate\cite{lai1} and 
the needle shown in Fig.~\ref{fig1}(b1). 
The shared linearity might however be a coincidence since  
 the proportionality constant $\Omega$ 
 seems to depend 
 strongly on the apex angle $\alpha$.
 By linear fits to the data, see dashed lines in Fig.\ref{fig:melt}(a), $\Omega$
 was extracted for all the studied needles.
 The result of these fits, shown in  Fig.~\ref{fig:melt}(b) and table~\ref{tab:tab1}, 
 reveals that $\Omega(\alpha)$
 increases significantly when the apex angle $\alpha$ is increased from
 26$^\circ$ to  42$^\circ$. 
 This implies that for a given radius $R$ the reduced melting temperature $\epsilon(T_c)$
 may vary dramatically. Thus the melting of the spherical interface is 
 not controlled by the local thickness of the needle.
 This conclusion is further supported directly from the TEM images since the 
 melting front does not follow the contours of the thickness fringes.

   \section{Discussion}
   We now turn to discuss the melting process from simple geometric considerations assuming
   that the needles have a perfect conical shape.
   The cap ratio $\mathcal{CR}(\alpha)=h(T)/R(T)$ relates, in this case, to the apex angle $\alpha$
   by 
   \begin{equation}\label{eq:contactangle}
   \mathcal{CR}(\alpha)=\frac{h(T)}{R(T)}=1-\sin(\alpha/2+\theta),
   \end{equation}
   where $\theta$ is the contact angle between the carbon coat and the solid-liquid
   interface.
The fact that the cap ratio $\mathcal{CR}(\alpha)$ is approximately temperature independent
for a fixed apex angle $\alpha$ implies 
that $\theta$ remains constant as the melting front moves towards the thicker 
part of the needle. Assuming that we know 
$\alpha$ with a reasonable degree of precision ($\pm1.0^{\circ}$) one can thus derive the contact 
angle $\theta$ from a constant fit to the data as shown by dashed lines 
in Fig.~\ref{fig:fig3}. We find that $\theta<1.2^{\circ}$ is systematically small 
($\theta/\alpha\ll 1$) for all the studied needles, see table~\ref{tab:tab1}. In fact $\theta$ is smaller or comparable to the 
estimated error of $\alpha$, it is therefore reasonable to assume 
that $\theta\approx0$. The solid line in Fig.~\ref{fig:fig3}(b) shows 
$\mathcal{CR}(\alpha)$ calculated from Eq.~\ref{eq:contactangle} with $\theta=0$.

Next we are going to discuss the ratio between the molten volume and the solid-liquid 
interface area.
For a given temperature $T_i<T<T_m$ we observed a spherical solid-liquid interface with radius $R$.
Now, if the temperature increases to $T^\prime=T+dT$, a part of the solid volume is going to melt. 
This leads to a larger solid-liquid interface with radius $R^{\prime}=R+dR$.
The molten volume $V_m$  therefore increases to $V_m^\prime=V_m+dV_m$ where 
\begin{equation}\label{eq:vol}
dV_m=\left\{\frac{\cos^3}{\tan}\left(\frac{\alpha}{2}\right)
-\mathcal{CR}^2(\alpha)\left[3-\mathcal{CR}(\alpha)\right]\right\}\pi R^2dR.
\end{equation}
Naturally, the solid-liquid interface area $S_{cap}$ increases as well to $S_{cap}^\prime=S_{cap}+dS_{cap}$ 
with
\begin{equation}\label{eq:sur1}
dS_{cap}=\mathcal{CR}(\alpha)4\pi R dR
\end{equation}
and the liquid-coat area $S_{lc}$ increases with
\begin{equation}\label{eq:sur2}
dS_{lc}=\sqrt{1+\frac{\cos^2}{\tan^2}\left(\frac{\alpha}{2}\right)} 
2\pi R dR.
\end{equation}
We are now going to assume that the condition for melting is that the energy gained upon 
 melting a solid volume $dV_m$ has to be balanced with the increased interface area $dS_{cap}+dS_{lc}$.
 Hence 
 \begin{equation}\label{eq:melcon}
 \Lambda(T_c) dV_m= \gamma_{sl} dS_{cap}+(\gamma_{sc}-\gamma_{lc}) dS_{lc}
 \end{equation}
 where $\Lambda(T_c)\approx L_m(T_m-T_c)/T_m=L_m\epsilon(T_c)$ is the latent heat per unit volume, $L_m$ is the latent heat of fusion,  and $\gamma_{sl}$ ($\gamma_{sc}$, $\gamma_{lc}$) is the solid-liquid (solid-coat, liquid-coat)
 interface energy.
 
 \begin{figure}
\includegraphics[width=0.45\textwidth]{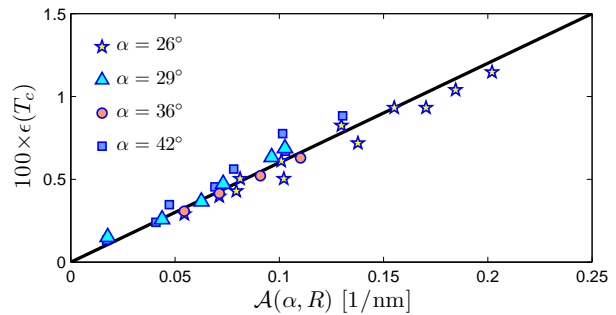}
\caption{\label{fig:fig4} Reduced melting temperature $\epsilon(T_c)$ as a function of surface area to molten volume ratio $\mathcal{A}(\alpha,R)=\mathcal{F}(\alpha)/R$ where $\mathcal{F}(\alpha)$ is given in the text.}
\end{figure}

 Now since the contact angle $\theta$ between the carbon coat and the solid-liquid interface is 
 approximately zero we can assume, via Young's relation ($\gamma_{lc}=\gamma_{sc}+\gamma_{sl}\cos(\pi+\theta)$), that $\gamma_{sl}\approx\gamma_{sc}-\gamma_{lc}$.
 
From Eqs.~(\ref{eq:vol}), (\ref{eq:sur1}), (\ref{eq:sur2}), and  (\ref{eq:melcon}) we now have
\begin{equation}
\epsilon(T_c)=\frac{\gamma_{sl}\mathcal{F}(\alpha)}{L_m R}
\end{equation}
where  $\mathcal{A}(\alpha,R)=\mathcal{F}(\alpha)/R$ is the surface area to molten volume ratio
and
\begin{equation}\label{eq:F}
\mathcal{F}(\alpha)=\frac{4\mathcal{CR}(\alpha)+2\sqrt{1+\frac{\cos^2}{\tan^2}\left(\frac{\alpha}{2}\right)}}
{\frac{\cos^3}{\tan}\left(\frac{\alpha}{2}\right)
-\mathcal{CR}^2(\alpha)\left[3-\mathcal{CR}(\alpha)\right]}
\end{equation}
is a geometrical factor depending only on the apex angle $\alpha$.
In table~\ref{tab:tab1}, $\mathcal{F}(\alpha)$ is evaluated for each of the needles.

The depressed melting temperature $\epsilon(T_c)$ for a needle with fixed apex 
angle $\alpha$ is therefore, in this simple model, 
proportional to the inverse radius $1/R$ as observed by the experiment.
Furthermore the model predicts that the proportionality constant 
$\Omega(\alpha)=\mathcal{F}(\alpha)\gamma_{sl}/L_m$ where $L_m\approx867$ MJ/m$^3$ (Ref.~\onlinecite{hand}).
To compare the model with the data, the only adjustable
parameter is the solid-liquid interface energy $\gamma_{sl}$. 
The solid line in Fig.~\ref{fig:melt}(b) shows $\Omega(\alpha)$ as predicted 
by the model with $\gamma_{sl}=55$ mJ/m$^2$.  

The reasonable agreement between the model and the data indicates that 
the $\alpha$-dependence of the proportionality constant $\Omega(\alpha)$  
should be understood 
directly from the  
surface area to molten volume ratio $\mathcal{A}(\alpha)$.
To visualize this point, we show in Fig.~\ref{fig:fig4} the depressed melting 
temperature $\epsilon(T_c)$ as a function of 
$\mathcal{A}(\alpha,R)$
for all the studied needles.
Now all the data points collapse onto a single line with the slope $\gamma_{sl}/L_m$.
Our results therefore strongly suggest that the melting process is governed by
the surface area to molten volume ratio and not by the local size of the system. 
Notice that while this might seem as an obvious conclusion from a
theoretical point of view it is not trivial to demonstrate it experimentally.

\begin{table}
\caption{\label{tab:tab1} Cap ratio $\mathcal{CR}(\alpha)$, contact angle $\theta$, proportionality constant $\Omega(\alpha)$, and the geometric factor 
$\mathcal{F}(\alpha)$ for the four studied needles. The cap ratio $\mathcal{CR}(\alpha)$ and the    proportionality constant $\Omega(\alpha)$ were derived from  fits to the data shown by dashed lines in Fig.~\ref{fig:fig3} and Fig.~\ref{fig:melt}(a). Knowing $\mathcal{CR}(\alpha)$ the contact 
angle $\theta$ can be calculated from Eq.~(\ref{eq:contactangle}) while $\mathcal{F}(\alpha)$ can be evaluated 
directly from Eq.~(\ref{eq:F}). Notice that for one needle $\alpha=36^{\circ}$ the contact angle $\theta$
has a small negative value which is an unphysical result. This negative 
sign is related to the experimental uncertainty involved in determining $\theta$.  }
\begin{ruledtabular}
\begin{tabular}{ccccc}
$\alpha$ & 26$^{\circ}$& 29$^{\circ}$&36$^{\circ}$&42$^{\circ}$  \\
  \hline
 $\mathcal{CR}(\alpha)$ & 0.761(4)& 0.745(7) &0.709(3)&0.609(8) \\  
$\Omega(\alpha)$ [nm]&0.23(1)&0.24(3)&0.281(7)&0.35(2)\\
$\theta/\alpha$ & 0.032& 0.019 &-0.030&0.048  \\ 
$\mathcal{F}(\alpha)$ & 4.08& 4.34&4.91&5.36\\
\end{tabular}
\end{ruledtabular}
\end{table}
By varying the apex angle $\alpha$ we succeeded to change the surface area to molten volume ratio 
$\mathcal{A}(\alpha,R)$
in a systematic fashion. This is 
shown in Fig.~\ref{fig:fig3} --
as the apex angle $\alpha$ increases the cap ratio $\mathcal{CR}(\alpha)$ decreases, which 
in turn leads to a corresponding increase of the 
surface area to molten volume ratio $\mathcal{A}(\alpha,R)\sim\mathcal{F}(\alpha)$, see table~\ref{tab:tab1}.
Thus we managed to vary  $\mathcal{A}(\alpha,R)$ without necessarily
 to  change the radius $R$
as opposed to the case of spherical particles.
The advantage of the needle system is therefore that we could show that
the  depressed melting temperature
$\epsilon(T_c)$ scales directly with the surface area to molten volume ratio
$\mathcal{A}(\alpha,R)$ by varying both $\alpha$ and $R$.

We now return to the discussion of the finite size melting of metal particles on a 
substrate.
The literature contains examples of both linear~\cite{buffat,lai} and non-linear~\cite{dippel}  relationships
between the depressed melting temperature $\epsilon(T_c)$ and the inverse radius.
Our results suggest that a linear relation is observed only if the surface area to molten volume 
ratio scales with $1/R$. This  is of course the case for free standing spherical particles.
However, for particles deposited on a substrate, that might not always be the case.
Recently J. Murai {\it et al.} \cite{murai} showed 
that the contact angle of Sn and Bi particles, 
supported by a substrate, is size dependent for $R<20$~nm. This indicates 
a more complicated surface area to molten volume ratio.

We close the discussion by commenting on the extracted value of the solid-liquid 
interface energy  $\gamma_{sl}=55$~mJ/m$^2$ of aluminium.
Previous experiments on aluminium particles deposited on a substrate suggested
values of $\gamma_{sl}$ that is a factor of two larger. \cite{turnbull,falken,lai1}
These values should, however, be compared cautiously because {\it (i)} 
different models have been used and {\it (ii)} 
the shape of metal particles on a substrate may not be
perfectly spherical due to the presence of the substrate {\it i.e.} the 
exact surface area to molten volume ratio is typically not specified. 
Finally, {\it (iii)} we stress that in our model we assumed that $\gamma_{sl}\approx\gamma_{sc}-\gamma_{lc}$.
However, if we instead assume $\gamma_{sl}\gg\gamma_{sc}-\gamma_{lc}$, {\it i.e.}
 $\gamma_{sc}\approx\gamma_{lc}$, the model would capture the data
equally well albeit $\gamma_{sl}\approx180$~mJ/m$^2$.
The exact value of $\gamma_{sl}$ therefore rely strongly on the 
relatively unknown interface energies of the carbon coat.
\section{Conclusion}
In summary, we have reported on the melting of aluminium when confined to 
nano-sized needles. The melting initiates at the needle tip and an equilibrium 
solid-liquid interface could be followed as the melting front moves towards
the thicker part of the needle with increasing temperatures.
The solid-liquid interface has a spherical topology {\it i.e.} the interface forms 
a spherical cap. For a needle with fixed apex angle
the melting is found to be proportional to the inverse radius of the spherical cap.
However, by varying the apex angle we showed that the proportionality constant depends 
on the ratio between the solid-liquid interface area and the molten volume.
This lead us to conclude that the {\it finite size} melting of aluminium is controlled merely
by the surface area to molten volume ratio
rather than the actual local size of the system.

\section{Acknowledgments}
This work was financially supported by the Japan Society for the
Promotion of Science under Grants-in-Aid for Scientific Research CA
(Contact N$^{\circ}$ 14205092), and the Danish Natural Science Research
Council. J. C. would like to thank the Association of International Education Japan (AIEJ), Julie Marie Vinter Hansens Fond, Nordea Danmark Fond, Julie Damms Fond and Vordingborg Gynmasium's fond for financial support during this work.

\end{document}